\documentclass[12pt]{iopart}

\usepackage{iopams}
\usepackage{graphicx}
\usepackage{color}

\newcommand{\ket}[1]{\left| #1 \right>} 
\newcommand{\bra}[1]{\left< #1 \right|} 

\begin{document}

\title[]{Optical properties of an inhomogeneously broadened multilevel V-system in the weak and strong probe regimes}

\author{Paramjit Kaur$^{1,2}$, Vineet Bharti$^3$ and Ajay Wasan$^1$}
\address{{$^1$Department of Physics, Indian Institute of Technology, Roorkee 247667, India}}
\address{{$^2$Department of Physics, Guru Nanak Dev University, Amritsar 143 005, India}}\address{{$^3$Department of Physics, Indian Institute of Science, Bangalore 560 012, India}}
\ead{pjit12@gmail.com}

\begin{abstract}
We present a theoretical model, using density matrix approach, to study the effect of weak as well as strong probe field on the optical properties of an inhomogeneously broadened multilevel V-system of the $^{87}$Rb D2 line. We consider the case of stationary as well as moving atoms and perform thermal averaging at room temperature. The presence of multiple excited states results in asymmetric absorption and dispersion profiles. In the weak probe regime, we observe the partial transparency window due to the  constructive interference occurs between transition pathways at the line center.  In a room temperature vapour, we obtain an increased linewidth of the transparency window and steep positive dispersion. For a strong probe regime, the transparency window with normal dispersion switches to enhanced absorption with anomalous dispersion at the line center. Here, we show how the electromagnetically induced transparency (EIT) depends on the polarizations of the applied fields. We also discuss the transient behaviour of our system which agrees well with the corresponding absorption and dispersion profiles. This study may help to understand optical switching and controllability of group velocity.
\end{abstract}


\section{Introduction}
Over the last two decades, the study of modification and control over the optical properties of an atomic system has attracted a great attention. In most of these studies, strong resonant field are used to control the optical properties of an atomic medium. Consequently, this has led to many important phenomena such as electromagnetically induced transparency (EIT) \cite{BIH91,HSE97}, lasing without inversion \cite{ZLN95,BKS11} and coherent population trapping (CPT) \cite{AE96}. But in recent years, the phenomenon of EIT has been intensively studied theoretically as well as experimentally in a three-level atomic system that can be one of the configuration among  $\Lambda$, $\Xi$ and ${\rm V}$ systems. EIT is a phenomenon, in which the absorption and dispersion properties of probe field across a transition modify if a strong control field is applied on another transition that shared one of the levels with first transition \cite{BIH91,HSE97,FIM05}. Many interesting and potential applications have already been shown using EIT, such as precision magnetometers \cite{BR07},  probe amplification \cite{MEA96}, optical switching \cite{BHB09,ADD11}, quantum memories \cite{FIM05}, high-resolution spectroscopy \cite{KPW05},  enhancement of second- and third-order nonlinear processes \cite{HFI90,ZAX08}, reduction of quantum noise \cite{GW94}, quantum correlations \cite {FM94,AGA93}, storage of light \cite{LDB01} and enhancement of electro-optical effect \cite{MBW08}. EIT has also a remarkable feature to control the group velocity of light in a material medium. The group velocity of light can be made small \cite{HHD99} or large than the speed of light in vacuum or even it becomes negative \cite{WKD2000,ABL99}.

In this work, we have considered a V-type EIT system which has been studied to lesser extent compared to $\Lambda$- and $\Xi$- systems. It is well known that in the three-level $\Lambda$-system, the strong control field optically pumps the atoms in the lower level of the probe transition. In the three-level $\Xi$-system, the strong control field is applied on the upper two unpopulated levels. While in the three-level V-system, two upper levels are connected by the probe and control fields to the common ground state that is initially fully populated.
In a V-system, the coherence dephasing rate created between the upper levels is much higher than the $\Lambda$- and $\Xi$- systems. EIT in the V-type system can be used for amplification or lasing without inversion \cite{MPC98}, and laser frequency stabilization \cite{RPM07}. To observe  V-type EIT, many experiments are performed in atomic vapours \cite{WPF98,ZWX02,VBA07,MDH13}. EIT has been observed in Na$_2$ molecule too \cite{LKA13}. More recently, Kumar \emph{et al.} have theoretically studied the additional one-photon coherence-induced transparency in a Doppler-broadened V-type system \cite{KMS13}. They have shown that there is a feasibility of attaining nearly perfect transparency using an additional photon in a Doppler broadened V-type system.

It is well known that most of the EIT studies are performed in alkali metals atomic vapour considering a three-level system in it. But in reality, there are complicated hyperfine levels associated with these atoms. We cannot ignore these near by levels as the applied field interact with them too. So the transitions among these degenerate magnetic sublevels have to be considered.
It has already been studied that the presence of these hyperfine levels affects the absorption profiles significantly in the various configurations of EIT \cite{BHA01,MSL11,BW12,BW13,KBW14}.

In this paper, we discuss the EIT phenomena in V-type system that involves only the discrete levels. But many studies of EIT have already been done by considering the upper excited states as a structured continuum (i.e. presence of an auto-ionizing (AI) state) \cite{SJP94, AMJ06, KVW12, TVW14}. van Enk \emph{et al.} have first discussed the model of a $\Lambda$- system in which upper excited states are replaced by a flat continuum \cite{SJP94}. It has been shown that optical properties of an $\Lambda$-atomic medium change significantly in which upper excited state is replaced by a structured continuum including AI resonances \cite{AMJ06}. Quite recently, Dinh \emph{et al.} have studied that the occurrence of an additional EIT window is due to the presence of second AI state \cite{TVW14}.

In many experiments, studies of EIT have focused largely on the weak probe limit. But departing from this weak limit,
leads to a deterioration of EIT and can result in electromagnetic induced absorption (EIA) that has been verified both theoretically as well as experimentally \cite{WG98}. The strong probe has mostly used with $\Xi$-system at room temperature. This strong probe causes a reduction in the absorption for stationary atoms and splitting of the EIT resonance occurs at the line centre for moving atoms \cite{PN08}. Recently, Pandey \emph{et al.} have studied a similar system and have shown that EIT lineshapes change when the probe field becomes strong \cite{PKN11}.

In present work, we investigate the effect of the weak as well as strong probe fields on the optical properties of multi-level V-type EIT system. We present absorption and dispersion profiles for stationary as well as moving atoms by performing thermal averaging at the room temperature. Initially, we consider a six-level system is formed by applying  $\sigma^{+}_{p}-$polarized probe and $\pi_{c}-$polarized control fields that are in the $ D2 $ line of $^{87}$Rb atom. This six-level system is chosen in such a way that there is no optical pumping into the extreme ground state of the atom. While studying the polarization effect on the EIT window, we have considered the different configuration of V-type system in the taken $ D2 $ line. In the strong probe field regime, we can also consider a five-level system which is formed by $\sigma^{+}_{p}$ probe and $\sigma^{-}_{c}$ control polarized fields. Further, we have solved the time-dependent density matrix equations to investigate the transient behavior of a V-type EIT system.

The arrangement of the paper is as follows. In section 2, a theoretical model for our six- and five-level systems is presented by using density matrix approach. In section 3, the numerical results are discussed and are compared with the similar three-level system. Finally, section 4 provides a conclusion of this theoretical work.

\section{Theoretical model}
\begin{figure}[tbh!]
\begin{center}
\includegraphics[width=0.8\columnwidth]{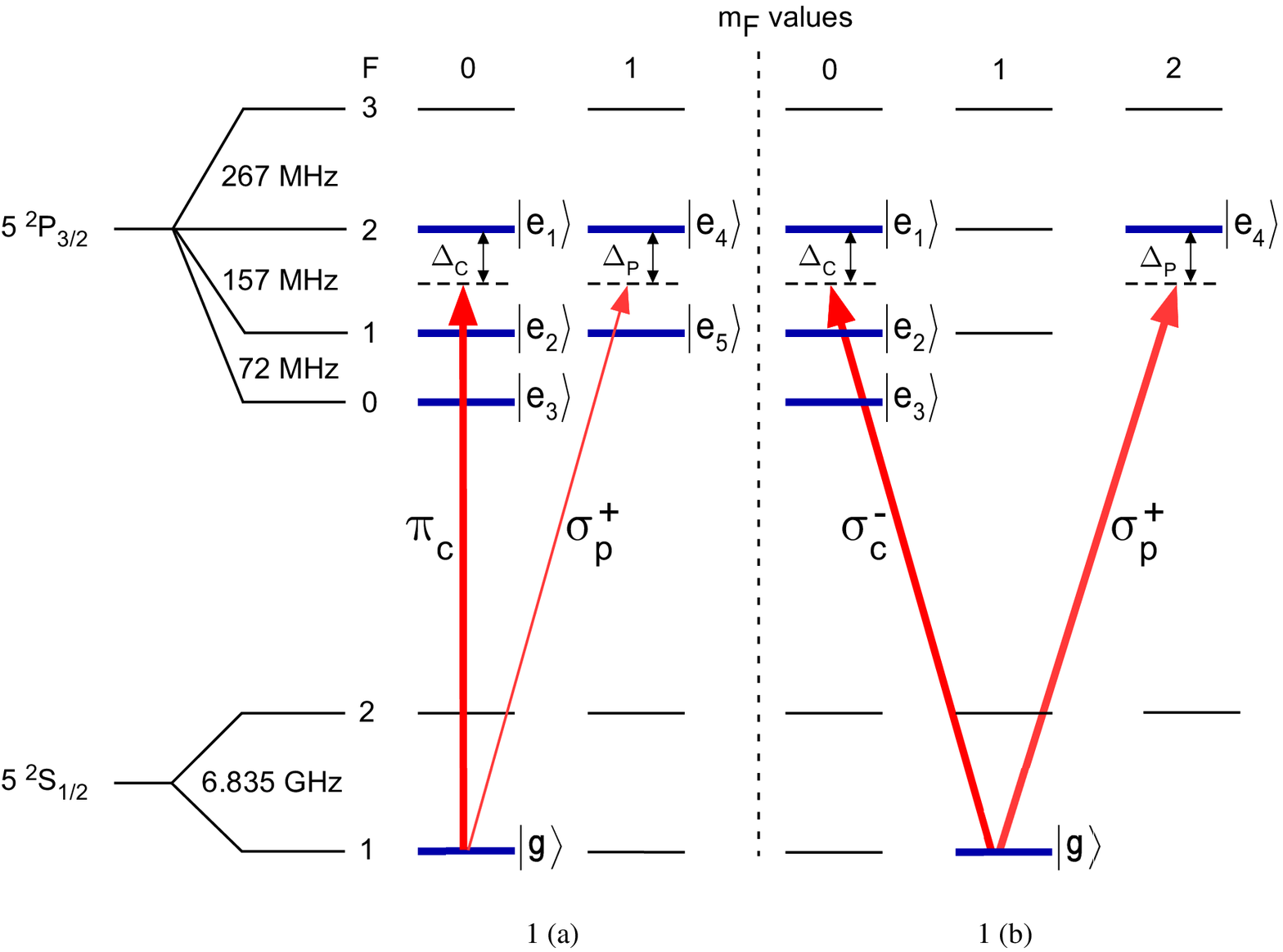}
\caption{\label{fig1} A multilevel ${\rm V}$-system of $^{87}$Rb-D2 line with only the considered m$_{F}$ values. In the six-level system, $\ket g$ is the ground state and  $\ket {e_{1}}$,  $\ket {e_{2}}$, $\ket {e_{3}}$, $\ket {e_{4}}$, $\ket {e_{5}}$ are the excited states. In the five-level system, only $\ket g$, $\ket {e_{1}}$, $\ket {e_{2}}$, $\ket {e_{3}}$, $\ket {e_{4}}$ states are present. The three-level approximation involves only $\ket g$, $\ket {e_{1}}$, $\ket {e_{4}}$ states. }
\end{center}
\end{figure}

We consider a six-level V-type system in $^{87}$Rb interacting with two electromagnetic fields as shown in figure~\ref{fig1}. The six-level system involves one hyperfine level of a ground state ${\rm 5^{2}S}_{1/2}$ and five hyperfine levels of the excited state ${\rm 5^{2}P}_{3/2}$. The $\pi_{c}-$polarized control field couples ground state $\ket g \equiv \ket {F=1, m_{F} =0}$ with excited states $\ket {e_{1}} \equiv \ket {F=2, m_{F} =0}$, $\ket {e_{2}} \equiv \ket {F=1,\\ m_{F} =0}$, $\ket {e_{3}} \equiv \ket {F=0, m_{F} =0}$. It may be noted that the dipole moment for $\ket g \leftrightarrow \ket {e_{2}}$ transition is zero, so the overall contribution of this transition is zero. Further, $\sigma^{+}_{p}-$polarized probe field couples ground state $\ket g$ with the excited states $ \ket {e_{4}} \equiv \ket {F=2, m_{F} =1}$ and $ \ket {e_{5}} \equiv \ket {F=1, m_{F} =1}$. The control and probe fields with frequency $\omega_{c}$ and $\omega_{p}$ are detuned from the atomic transition $\ket g \rightarrow \ket {e_{1}}$ and $\ket g \rightarrow \ket {e_{4}}$   by $\Delta_{c}=\omega_{e_{1}g}-\omega_{c}$ and  $\Delta_{p}=\omega_{e_{4}g}-\omega_{p}$ , respectively. Here  $\omega_{jk}=(E_{j}-E_{k})/\hbar$ is defined as the atomic transition frequency between levels $j$ and $k$ ($j>k$), and $E_{j}$ is the energy of the unperturbed atomic state $\ket {j}$.  Our six-level system can be reduced to a simple three-level system if only  $\ket g$,  $\ket {e_{1}}$, $\ket {e_{4}}$ states are present.

The Hamiltonian of the six-level system after carrying out the rotating-wave approximation is written as
\begin{eqnarray}
\hat{H}&=& -\hbar\left(\displaystyle\sum_{j=1}^3(\Delta_{c}+\omega_{e_{j}e_{1}})\ket {e_{j}}\bra{e_{j}}+\displaystyle\sum_{k=4}^5(\Delta_{p}+\omega_{e_{k}e_{4}})\ket {e_{k}}\bra{e_{k}}\right)\nonumber\\&+&\frac{\hbar}{2}\left(\displaystyle\sum_{j=1}^3\Omega_{cge_{j}}\ket {g}\bra{e_{j}}+\displaystyle\sum_{k=4}^5\Omega_{pge_{k}}\ket {g}\bra{e_{k}}+h.c.\right),
\end{eqnarray}where \textit{h.c.} is the complex conjugate of the preceding off-diagonal terms. The time evolution of the system in the semi-classical theory is governed by \textit{Liouville equation} \cite{ABR10}:
\begin{equation}
 \dot{\rho}= -\frac{i}{\hbar}[H,\rho] - \frac{1}{2}\{\Gamma,\rho\}.
\end{equation}Here $\Gamma$ incorporates the decay rate of each state. By substituting expression (1) into Eq. (2), we obtain the following set of density matrix equations:

\begin{eqnarray}
\dot{\rho}_{gg}&=&-\frac{\rho_{gg}}{\tau_{d}}+\frac{1}{\tau_{d}}+
\frac{b_{e_{1}}}{3}\Gamma_{e_{1}}\rho_{e_{1}e_{1}}+
\frac{b_{e_{3}}}{3}\Gamma_{e_{3}}\rho_{e_{3}e_{3}}+
\frac{b_{e_{4}}}{2}\Gamma_{e_{4}}\rho_{e_{4}e_{4}}+
\frac{b_{e_{5}}}{2}\Gamma_{e_{5}}\rho_{e_{5}e_{5}}\nonumber\\&+&
\displaystyle\sum_{j=1}^3\frac{i}{2}(\Omega^{*}_{cge_{j}}\rho_{ge_{j}} - \Omega_{cge_{j}}\rho_{e_{j}g})+\displaystyle\sum_{k=4}^5\frac{i}{2}(\Omega^{*}_{pge_{k}}\rho_{ge_{k}}-
\Omega_{pge_{k}}\rho_{e_{k}g}),\nonumber\\
\dot{\rho}_{e_{j}e_{j}}&=&-\Gamma_{e_{j}}\rho_{e_{j}e_{j}}+
\frac{i}{2}(\Omega_{cge_{j}}\rho_{e_{j}g}-\Omega^{*}_{cge_{j}}\rho_{ge_{j}}),\nonumber\\
\dot{\rho}_{e_{k}e_{k}}&=&-\Gamma_{e_{k}}\rho_{e_{k}e_{k}}+\frac{i}{2}(\Omega_{pge_{k}}\rho_{e_{k}g}-\Omega^{*}_{pge_{
k}}\rho_{ge_{k}}),\nonumber\\
\dot{\rho}_{ge_{j}}&=&\left(-\frac{\Gamma_{e_{j}}}{2}+i({\Delta_{c}}+
\omega_{e_{j}e_{1}})\right)\rho_{ge_{j}}+\frac{i}{2}\Omega_{cge_{j}}(\rho_{gg}-\rho_{e_{j}e_{j}})-\displaystyle\sum_{j\neq l,l=1}^3\frac{i}{2}(\Omega_{cge_{l}}\rho_{e_{l}e_{j}})
\nonumber\\&-&\displaystyle\sum_{k=4}^5\frac{i}{2}(\Omega_{pge_{k}}\rho_{e_{k}e_{j}}),\nonumber\\
\dot{\rho}_{ge_{k}}&=&\left(-\frac{\Gamma_{e_{k}}}{2}+i(\Delta_{p}+
\omega_{e_{k}e_{4}})\right)\rho_{ge_{k}}+\frac{i}{2}\Omega_{pge_{k}}(\rho_{gg}-\rho_{e_{k}e_{k}})
-\displaystyle\sum_{j=1}^3\frac{i}{2}(\Omega_{cge_{j}}\rho_{e_{j}e_{k}})\nonumber\\&-&\displaystyle\sum_{k\neq m,m=4}^5\frac{i}{2}(\Omega_{pge_{m}}\rho_{e_{m}e_{k}}),\nonumber\\
\dot{\rho}_{e_{k}e_{j}}&=&\left(-\frac{(\Gamma_{e_{k}}+\Gamma_{e_{j}})}{2}-i({\Delta_{p}-\Delta_{c}}-
\omega_{e_{j}e_{k}})\right)\rho_{e_{k}e_{j}}+
\frac{i}{2}(\Omega_{cge_{j}}\rho_{e_{k}g}-\Omega^{*}_{pge_{k}}\rho_{ge_{j}}),\nonumber\\
\dot{\rho}_{e_{j}e_{l}}&=&\left(-\frac{(\Gamma_{e_{j}}+\Gamma_{e_{l}})}{2}+
i\omega_{e_{j}e_{l}}\right)\rho_{e_{j}e_{l}}+\frac{i}{2}(\Omega_{cge_{l}}\rho_{e_{j}g}-\Omega^{*}_{cge_{j}}\rho_{ge_{l}}),\nonumber\\
\dot{\rho}_{e_{k}e_{m}}&=&\left(-\frac{(\Gamma_{e_{k}}+\Gamma_{e_{m}})}{2}+
i\omega_{e_{k}e_{m}}\right)\rho_{e_{k}e_{m}}+\frac{i}{2}(\Omega_{pge_{m}}\rho_{e_{k}g}-\Omega^{*}_{pge_{k}}\rho_{ge_{m}}).\nonumber\\
\end{eqnarray}The $\Omega_{cge_{j}}$ is the Rabi frequency of the control field for the transition $\ket g\leftrightarrow \ket {e_{j}}$ with \emph{j} = 1, 2, 3 and $\Omega_{pge_{k}}$ is the Rabi frequency of the probe field for the transition $\ket g  \leftrightarrow \ket {e_{k}} $ with \emph{k} = 4, 5. In the above density matrix equations, $\rho_{jj}$ denotes the population of state $\ket j$ and $\rho_{jk}$ denotes the coherence between states $\ket j$ and $\ket k$. Complex conjugates of coherence and Rabi frequencies are given by $\rho_{jk} = \rho^{*}_{jk}$, $ \Omega_{cgj} = \Omega^{*}_{cgj}$ and $ \Omega_{pgk} = \Omega^{*}_{pgk}$. The branching ratio of the $j^{th}$ level is given by $b_{j}$. The spontaneous decay rate of the excited states $\ket {e_{1}}$, $\ket {e_{2}}$, $\ket {e_{3}}$, $\ket {e_{4}}$, $\ket {e_{5}}$ are $\Gamma_{e_{1}} = \Gamma_{e_{2}}=\Gamma_{e_{3}} = \Gamma_{e_{4}} =\Gamma_{e_{5}}= 2\pi\times6.1$ MHz \cite{SAS11}. We have also introduced  the ground-state decay rate because there is finite interaction time $\tau_{d}$ of the atom with an electromagnetic (e.m.) field. For our numerical calculations, we assume that the loss rate of atoms is much smaller than the radiative decay rates of the excited states, $\tau^{-1}_{d} \ll \Gamma_{e_{1},e_{2},e_{3},e_{4},e_{5}}$. The radiative decay rate of the excited states are 26.1 ns, so we assume $\tau_{d} \approx 300\mu s $ which corresponds to the interaction time with e.m. fields of a few millimeters in diameter for all temperature \cite{MSL11,BW12}. We also assume that the rate at which new atoms are entering the interaction region is equal to the rate at which atoms are leaving the interaction region. This rate is included by the term ${1}/{\tau_{d}}$. The matrix elements of dipole moments and corresponding Rabi frequencies of all the transitions in our considered V-system are calculated by using the Clebsch-Gordan coefficient for hyperfine levels \cite{ABR10,BW12}.

In this work, we mainly focus on response of the medium to the probe field which is determined by coherence $(\rho_{ge_{4}})$ between levels $\ket g$ and $\ket {e_{4}}$. The susceptibility $\chi$ is a response function to the applied  e.m. field and is calculated as:
\begin{eqnarray}
\chi=\chi^\prime+i\chi^{\prime \prime}= {\frac{N {\mid \mu_{ge_{4}} \mid}^2}{\hbar \epsilon_{0}\Omega_{p}}}\rho_{ge_{4}}\;,
\end{eqnarray}

where $\chi^\prime$ and $\chi^{\prime \prime}$ are the real and imaginary parts of the susceptibility. \textit{N }is atomic number density in the medium and $\mu_{ge_{4}}$ is the dipole matrix element between the levels $\ket g$ and $\ket {e_{4}}$. The knowledge of susceptibility gives complete description of the dispersion and absorption of the probe field. Therefore, the dispersion and absorption of the probe field is proportional to $Re(\rho_{ge_{4}})$ and $Im(\rho_{ge_{4}})$, respectively.

The group velocity of the probe field is given by:
\begin{eqnarray}
v_{g}=\frac{c}{n_{g}}=\frac{c}{1+\frac{\chi^\prime}{2}+\frac{\omega_{p}}{2}[\frac{\partial\chi^\prime}{\partial \omega_{p}}]}\;,
\end{eqnarray}here \textit{c} is the speed of light and $n_{g}$ is the group index. According to eq. (5), the slope of the probe dispersion determines the group velocity of the probe field. The slow group velocity or subluminal propagation of light occurs in optical media with normal dispersion whereas superluminal propagation of light occurs with anomalous dispersion [21].

The $\pi_{c}$ control field can excite the atoms from various ground state sub-levels. But the probability of spontaneous emission to $m_{F}=0$ sub-level is more. Therefore in our six-level system we have chosen $m_{F}=0$ sub-level as $\ket {g}$ state. The population conservation equation is:
$\rho_{gg}+\rho_{e_{1}e_{1}}+\rho_{e_{2}e_{2}}+\rho_{e_{3}e_{3}}+\rho_{e_{4}e_{4}}+\rho_{e_{5}e_{5}} = 1$. For a weak probe field, we assume that initially all the population is in the ground state, i.e. $\rho_{gg}\approx1$ and $\rho_{e_{1}e_{1}},\,\rho_{e_{2}e_{2}},\,\rho_{e_{3}e_{3}},\,\rho_{e_{4}e_{4}},\,\rho_{e_{5}e_{5}}\approx 0$. But as the Rabi frequency of the control field increases, it causes a distribution of population between the states $\ket {g}$, $\ket {e_{1}}$, $\ket {e_{2}}$ and $\ket {e_{3}}$. So, our assumption that initially all the population in the ground state, i.e. $\rho_{gg}\approx 1$ is no longer valid. The population terms $\rho_{gg},\,\rho_{e_{1}e_{1}},\,\rho_{e_{2}e_{2}},\,\rho_{e_{3}e_{3}}$  vary with increasing Rabi frequency of the control field and finally these population terms get saturated. In the numerical solution, the probe dispersion and absorption depend upon the population difference ($\rho_{gg}-\rho_{e_{j}e_{j}}$)  of the $\ket {g} \leftrightarrow \ket {e_{j}}$ transition which is driven by the control field. In this strong probe limit, there is population transfer between the levels which are driven by the control as well as the probe fields. The probe dispersion and absorption in this case depend upon the population differences ($\rho_{gg}-\rho_{e_{j}e_{j}}$) and ($\rho_{gg}-\rho_{e_{k}e_{k}}$). We have solved the above density-matrix equations numerically for a steady state solution ($\dot{\rho}=0$).

We consider the different configurations of V-system to demonstrate the dependence of EIT on the probe and control field polarizations. Our six-level system can be reduced to a five-level system if the state  $\ket {e_{5}}$ is absent. In this five-level system,  the ground state $\ket g \equiv \ket {F=1, m_{F} =1}$ is coupled with the excited states $\ket {e_{1}} \equiv \ket {F=2, m_{F} =0}$, $\ket {e_{2}} \equiv \ket {F=1, m_{F} =0}$, $\ket {e_{3}} \equiv \ket {F=0, m_{F} =0}$ by a $\sigma^{-}_{c}-$polarized control field and with excited state $\ket {e_{4}}$ $\equiv\ket {F=2, m_{F} =2}$ by a $\sigma^{+}_{p}-$polarized probe field, as shown in figure 1(b). The density matrix equations of a five-level system can be obtained from equations (3) by considering the Rabi frequency and decay rate of transition $\ket {e_{5}} \leftrightarrow \ket g  $ zero, as has been discussed in Appendix A. This five-level system is only applicable in the strong probe regime because in the weak probe regime, the strong $\sigma^{-}_{c}-$polarized control field optically pumps the atoms into the $m_{F}=-1$ sublevel of the ground state.

\section{Results and discussion}
In this section, we discuss the response of our considered V-system and compare it with a similar three-level system under weak and strong probe regimes. We have shown the absorption and dispersion profiles for stationary as well as for moving atoms. For the case of a moving atom, we perform thermal averaging at the room temperature to see the effect of all velocities groups on the probe absorption and dispersion profiles. Finally, we discuss the transient behaviour of our system.

\subsection{Optical susceptibility for stationary atom}
First we show the numerical results of the imaginary and real parts of  $\rho_{ge_{4}}$ versus the probe field detuning  ($\Delta_{p}$) for three- and six-level systems in the weak probe regime with different values of the control Rabi frequencies $(\Omega_{cge_{1}})$ in figure 2. The solid and dashed curves are for the six- and three- level systems, respectively.

In a three-level system, if the control field is considered in resonance with the corresponding transition, i.e. $\Delta_{c}=0$,   we do not obtain any transparency for $\Omega_{cge_{1}} < \Gamma_{e_{4}}$ as shown in figure 2(a). So in this case no Fano-type quantum interference takes place \cite{FIM05,BW12}. It means in a V-type EIT system the absorption profile does not split for quantum interference like in $\Lambda$- and $\Xi$- EIT systems. This can be explained by considering both the excited states in V-configuration. Though dipole transitions are forbidden in these excited states but still the coherence dephasing rate between them is non- zero and is much higher than $\Lambda$- and $\Xi$- configurations \cite{FSM95}. The corresponding dispersion curve has negative slope near the line center. The negative slope of the probe dispersion results in fast light propagation or superluminal light as expected from the expression given in Eq. 4.

For the case when $\Omega_{cge_{1}} > \Gamma_{e_{4}}$, the absorption splits into a doublet and shows a transparency window with non-vanishing probe absorption at zero probe detuning, shown in figure 2(b). These doublet peaks are known as Aulter-Townes doublet  that are  two symmetric dressed states created by the control field \cite{AUT55,ADS11,COR77}. The postion of these dressed states is given by the eigenvalues of the interaction Hamiltonian of three-level system which are $\pm \Omega_{cge_{1}}/2$. It means for the V-type EIT system, the probe absorption is enhanced rather than completely suppresses as in $\Lambda$- EIT system at the zero probe detuning. This enhanced probe absorption arises due to the fact that the interference between dressed states to a excited state for the V-type EIT system is constructive \cite{MEY01}, presented in Appendix B. But in case of $\Lambda$- EIT system, completely suppresses probe absorption is due to the destructive interference \cite{MEY01}. For this case ($\Omega_{cge_{1}} > \Gamma_{e_{4}}$), the dispersion curve becomes normal and shows a positive slope near zero probe detuning which leads to a subluminal light. As the $\Omega_{cge_{1}}$ varies, the separation between these dressed states, i.e. the linewidth of the transparency window increases symmetrically and slope of dispersion curve becomes more positive with respect to $\Delta_{p} = 0$ as shown in figure 2. Our calculations show as the Rabi frequency of control field increases, the subluminal phenomenon becomes more obvious in the weak probe regime. Beside this, our calculations show that the decay rates of the excited states significantly effect the EIT window. We obtain less transparency in V-system due to large decay rates of excited states.

In the case of a six-level system, modification of the absorption and dispersion profiles depend on the control field's Rabi frequency. We do not observe any deviation from a three-level system in the  absorption and dispersion profiles upto 10 MHz. If the control Rabi frequency increases further, an asymmetry starts appearing in both the absorption and dispersion profiles. With further increase above 15 MHz, the transparency window and dispersion profiles start shifting slightly toward the negative probe detuning, as shown in figure 2. This modification and shift of the absorption and dispersion profiles arise because our six-level system has multiple V-channels with different dipole moments for various transitions.

Now we extend our discussion in the strong-probe regime. We discuss the response of six- and five-level systems with varying strengths of probe field's Rabi frequency and are compared with three-level system as shown in figure 3. The red dash-dotted, blue solid and black dashed lines are for the six-, five- and three-level systems, respectively. In the three-level system when the probe Rabi frequency ($\Omega_{pge_{4}}$) is equal to 6 MHz, the probe absorption splits into the doublet peaks that gives a partial transparency window (figure 3(a)) and the corresponding dispersion curve has positive slope at the line center (figure 3(e)), similar to that of weak probe regime. With further increase in $\Omega_{pge_{4}}$, the linewidth of the particular absorption peak increases and its overall amplitude decreases that finally leads to the convolution of these two peaks emerges into a single enhanced absorption profile. The transparency window disappears at $\Omega_{pge_{4}}=24$ MHz. This modification is due to the saturation and power-broadening effect. Also, interference arises between the new absorption paths created by the probe field that were negligible when the probe field is weak \cite{PKN11}. The corresponding slope of the probe dispersion curve also reverses from positive to negative with increase in the Rabi frequency of the probe field. One clearly sees that our system shifts from subluminal to superluminal light as the probe field Rabi frequency increases.

In the case of a six-level system, the absorption and dispersion profiles are asymmetric which is expected due to the presence of multiple excited states. The position of absorption peaks changes significantly and the absorption and dispersion profiles shift toward the negative probe detuning. In this case, the whole spectrum of the absorption and  dispersion profiles seems to be red shifted.

The numerical analysis of a five-level system also show small transparency window with more asymmetry as compared to a six-level system. It is due to the fact that asymmetric nature of the absorption and dispersion profiles depends on the polarization of the applied fields. In a six-level system, the dipole moment for $\ket g \leftrightarrow \ket {e_{2}}$ transition is zero due to $ \pi_{c}-$polarized control field. For this reason, the nearest state is 229 MHz far away from the state $\ket {e_{1}}$. But in a five-level system, the control field is $\sigma^{-}_{c}-$polarized and the dipole moment correspond to transition $\ket g \leftrightarrow \ket {e_{2}}$  is non-zero so in this case, the nearest state is 157 MHz away from the state $\ket {e_{1}}$ . It may be noted that the overall absorption amplitude more in the strong probe regime as compare to the weak probe regime.

\subsection{Optical susceptibility for moving atom}
The analysis discussed in the previous section is for zero-velocity atoms. But we know that when the temperature is non-zero, the atoms move randomly due to their thermal energy. This thermal motion of atoms leads to a spreading of the absorbed frequency due to the Doppler effect which causes the broadening in the optical line profile.

It has already been studied that the overall transparency window for a three-level $\Lambda$ or $\Xi$-system shrinks after carrying out the thermal averaging of velocities at the room temperature vapour \cite{BW12, IKN08}. These two systems have also been studied with the presence of multiple excited states in the same condition. In this case, the transparency window for a $\Lambda$-system disappears totally  \cite{MSL11}, but for a $\Xi$-system it increases a bit more  \cite{BW12}. In this section, we are interested to see the effect of temperature on the optical properties of V-type system with multiple excited states. To see this effect, we solve susceptibility numerically as a function of temperature for a considered system.

We consider an atom moving with velocity $v$ along the direction of probe and control fields. The Doppler shifts
($\Delta_{D}$) for the probe and control fields are given by $\pm v/\lambda_{p}$ and $\pm v/\lambda_{c}$, where $\lambda_{p}$  and $\lambda_{c}$ are the wavelengths of the probe and control fields, respectively. The sign of the Doppler shift
depends upon the direction of the applied field, i.e. +ve sign is for an atom moving towards the applied field and $-$ve for an atom moving away from the applied fields. To make our calculations Doppler free, we consider that the probe and control fields are co-propagating. We perform the thermal averaging by using the Maxwell-Boltzmann velocity distribution, at temperature \textit{T}, which is given by
\begin{eqnarray}
f(v)=\frac{1}{u\sqrt{\pi}}exp\left(-\frac{v^{2}}{u^{2}}\right)\;.
\end{eqnarray}Here \textit{u} is most probable atomic velocity which is defined by
\begin{eqnarray}
u=\sqrt\frac{2kT}{m}\;,
\end{eqnarray}where \textit{k} is the Boltzmann constant, and \textit{m} the atomic mass. The most probable velocity for $^{87}$Rb corresponding to the room temperature (297 K) is 237.52 ms$^{-1}$. For our Doppler-broadened system,  the susceptibility of the probe field is averaged over all the velocities by using the Maxwell-Boltzmann distribution:
\begin{eqnarray}
\bar{\chi}=\int_{-\infty}^{\infty}\chi(v)f(v)dv
\end{eqnarray}The velocity-dependence of susceptibility is obtained from eq.(3) and (4) by replacing $\Delta_{p}\rightarrow\Delta_{p}\pm v/\lambda_{p}$ and $\Delta_{c}\rightarrow\Delta_{c}\pm v/\lambda_{c}$.

The results of thermal averaging of a six-level system in the weak probe limit with $\Delta_{c} = 0$ and $\Omega_{cge_{1}}=18$ MHz are shown in figure 4. Interestingly, we do not observe any narrowing in the  transparency window in our V-type EIT system as observed in  $\Lambda$- and  $\Xi$-systems \cite{MSL11,BW12,IKN08} after the thermal averaging. At the room temperature, we observe equal transparency windows for both the three- and six-level systems as shown in figure 4(a). The corresponding slope of the dispersion curve becomes positive at the zero probe detuning as shown in Figure 4(b).

In thermal averaging case, the broadening of transparency window can be understood by considering the effect of different atomic velocities on the probe absorption, as shown in Figure 5. It may be clearly seen that the position of the Aulter-Townes doublet is shifted right (left) from the center when atom is moving with velocity $\pm 5$ ms$^{-1}$, respectively. Hence, it does not fill the transparency region for a stationary atom as in $\Lambda$- and  $\Xi$-systems. This means that the overall transparency region becomes broader.

Now, we turn to the strong probe limit. The Doppler-broadened absorption and dispersion profiles of six-, five- and three-level systems in the strong probe  with  $\Omega_{cge_{1}}=18$ MHz and $\Omega_{pge_{4}}=24$ MHz are shown on the right-hand side of figure 4. Our results demonstrate that at room temperature, the degree of EIT decreases with increase in Rabi frequency of the probe field  $\Omega_{pge_{4}}$ in  three-, five- and six- level systems. There is an enhanced absorption at line center and two distinct transparency peaks arise on either side of line center as shown in figure 4(c). It is expected because the width of the individual peak of the Aulter-Townes doublet increases due to power broadening as we have already discussed. The corresponding slope of the probe dispersion curve is negative near $\Delta_{p}= 0$, as shown in figure 4(d).

The decrease in the degree of EIT for three-, five-, and six-level systems in the strong probe regime can be understood by considering the effect of velocity on the absorption profiles, as shown in figure 6. The solid lines are for stationary atoms ($v=0$), while the dashed and dash-dotted lines are for atoms moving with $v=+5$ and $-5$ ms$^{-1}$, respectively. It can be seen clearly that the symmetric transparency window for ($v=0$) becomes antisymmetric and overall absorption profile shift right (left) when velocity changes to $\pm$ 5 ms$^{-1}$. If we take the average of all velocities groups, the shifting of these peaks reduces the transparency region at the zero probe detuning. Also, there is a reduction in the probe absorption and dispersion at room temperature due to spreading of the absorption and dispersion over all the velocity groups.

\subsection{Transient evolution of the optical response}
In this section, we investigate the transient optical properties of a Doppler-broadened V type-system in the weak and strong probe regimes. These studies may be useful for optical information processing and controlling the group velocity of light from subluminal to superluminal.

We solve our six-level system's time-dependent density matrix equations to show the time evolution of Im$(\rho_{ge_{4}})$ and Re$(\rho_{ge_{4}})$. The time-dependent probe field absorption and dispersion profiles in weak and strong probe regimes are shown in figure 7. The solid and dashed curves are for the Im$(\rho_{ge_{4}})$ and Re$(\rho_{ge_{4}})$, respectively. First, we discuss the transient behaviour of our considered system in the weak probe regimes. Figure 7(a) shows that the probe absorption Im$(\rho_{ge_{4}})$ is zero at time $t=0$. But as the time increases, Im$(\rho_{ge_{4}})$ has a small oscillatory behaviour for few microseconds and finally it reaches steady state condition with non-vanishing probe absorption. The physical reason behind this non-vanishing probe absorption is the AC-stark effect caused by the strong control field that suppresses the partial probe absorption. The corresponding positive dispersion has an oscillating behaviour in small time interval and it finally reaches to steady state with time.

Now, we discuss the transient behaviour in the strong probe regime. Our results show that the probe absorption Im$(\rho_{ge_{4}})$ has a oscillating behaviour and finally reaches to very small absorption at the steady state. But, the corresponding dispersion is negative and reaches to the steady state very fast as shown in figure 7(b). Our numerical calculations for the transient response show that our six-level medium exhibits normal dispersion, i.e. subluminal light in the weak probe regime and anomalous dispersion, i.e. superluminal light in the strong probe regime. This transient evolution of optical response of the considered system can be explained by the probe absorption and dispersion profiles shown in figures 2 and 3.

\section{Summary and conclusion}
Our numerical calculations succeed in describing the optical response of the Doppler broadened V-system wih multiple excited states and help in analyzing the system in the weak and strong probe regimes. We have shown that in the weak probe regime, the probe field is always absorbed at the zero probe detuning for $\Omega_{cge_{1}} < \Gamma_{e_{4}}$. The partial transparency window is purely due to the AC-stark effect caused by the control field for $\Omega_{cge_{1}} > \Gamma_{e_{4}}$. Later case produces normal dispersion regions. But in the strong probe regime, our system shows anomalous dispersion region with enhanced absorption at the line center. Furthermore, one can clearly see that the regions for subluminal and superluminal light depend on the probe field's  Rabi frequency. Therefore, V-system can be used for manipulating the group velocities of light. Understanding of these studies will useful in  potential applications such as optical communications and networks.

We have also observed that induced transparency window arises similar to that of the weak probe limit when both fields are strong with the condition $\frac{\Omega_{pge_{4}}}{\Omega_{cge_{1}}} < 1$. But when we fix the control field's Rabi frequency with the condition $\frac{{\Omega_{pge_{4}}}}{\Omega_{cge_{1}}}\geq 1$, there is a reduction in the induced transparency window, i.e. the atomic coherence gives rise to enhanced absorption at the line center. This is due to the fact that the strong probe field causes additional population transfer between the states by creating new absorption paths. In addition to this, our theoretical model clearly indicates that the above disappeared transparency window can be recovered for a strong probe field by increasing the strength of the control Rabi frequency. Our numerical results show that the probe absorption and dispersion profiles become asymmetric due to the presence of multiple excited states. In these systems, the asymmetricity and amplitude of absorption profiles depend upon the polarization of probe and control fields. We also observe that the strength of EIT window strongly depends on the decay rates of the excited states.

We have also discussed the transient behaviour of a six-level system by solving the time-dependent density matrix equations. Our numerical results of the transient behaviour in steady state show that the transparency window with non-vanishing probe absorption for weak probe regimes and small enhanced absorption for strong probe regimes.

\section*{Acknowledgements}
PK is thankful to the Ministry of Human Resource Development (MHRD), India for the financial assistance. VB acknowledges financial support from a DS Kothari post-doctoral fellowship of the University Grants Commission, India.

\appendix
\section{Density matrix equations for a five-level system}
In this appendix, we discuss density matrix equations for a five-level system. These density matrix equations can be obtained from equations (3) by removing the contribution of the transition $\ket {e_{5}} \leftrightarrow \ket g$, in Figure 1(a). This will affect the population of the ground state $\rho_{gg}$ that will have to modify due to the different decay channels in the resulted five-level system Figure 1(b), so it can be written as

\begin{eqnarray}
\dot{\rho}_{gg}&=&-\frac{\rho_{gg}}{\tau_{d}}+\frac{1}{\tau_{d}}+
\frac{b_{e_{1}}}{3}\Gamma_{e_{1}}\rho_{e_{1}e_{1}}+
\frac{b_{e_{2}}}{3}\Gamma_{e_{2}}\rho_{e_{2}e_{2}}+
\frac{b_{e_{3}}}{3}\Gamma_{e_{3}}\rho_{e_{3}e_{3}}+
{b_{e_{4}}}\Gamma_{e_{4}}\rho_{e_{4}e_{4}}
\nonumber\\&+&\frac{i}{2}\displaystyle\sum_{j=1}^3(\Omega^{*}_{cge_{j}}\rho_{ge_{j}} - \Omega_{cge_{j}}\rho_{e_{j}g})+\frac{i}{2}\displaystyle\sum_{k=4}^5(\Omega^{*}_{pge_{k}}\rho_{ge_{k}}-
\Omega_{pge_{k}}\rho_{e_{k}g}).
\end{eqnarray}

\section{Dressed-state analysis}
In this appendix, we present semiclassical dressed-state picture for a three-level V-system. Our calculations show that for a case $\Omega_{cge_{1}} > \Gamma_{e_{4}}$, we observe the partial transparency window at the zero probe detuning. This can be explained by considering the dressed-state analysis of a three-level V-system. The control field couples the states $\ket g $ and $\ket {e_{1}}$ that create two dressed states  $\ket + $ and $\ket - $;

\begin{eqnarray}
\ket +=\frac{1}{\sqrt{2}}(\ket {e_{1}}+\ket g)\\
\ket -=\frac{1}{\sqrt{2}}(\ket {e_{1}}-\ket g).
\end{eqnarray}

The probe absorption is from one of these dressed states to an excited state $\ket {e_{4}}$, i.e. it has two excitation paths $\ket {e_{4}} \rightarrow \ket + $ and $\ket {e_{4}} \rightarrow \ket - $, which interfere with each other. The transition amplitude at the (undressed) resonant frequency $\omega_{e_{4}g}=(E_{e_{4}}-E_{g})/\hbar$, from the excited state $\ket {e_{4}}$ to the dressed states will be the sum of the contributions to states  $\ket + $ and $\ket - $, is given by
\begin{equation}
P\propto\left|\frac{\langle{ e_{4} | \textbf{d.E} | + \rangle}}{\Omega_{cge_{1}}}+\frac{\langle{ e_{4} | \textbf{d.E} | - \rangle}}{-\Omega_{cge_{1}}}\right|^{2}=\frac{{\Omega^{2}_{pge_{4}}}}{\Omega_{cge_{1}}}.
\end{equation}

The transition amplitude is not zero. It means for V-type EIT system, the interference between two excitation paths is constructive \cite {MEY01} and the probe absorption is enhanced at the zero probe detuning with transition amplitude given by eq.(B.3). Due to this reason, we observe only partial transparency window at zero probe detuning for $\Omega_{cge_{1}} > \Gamma_{e_{4}}$ case.

\newpage
\noappendix

\begin{figure}[tbh!]
\begin{center}
\includegraphics[width=0.8\columnwidth]{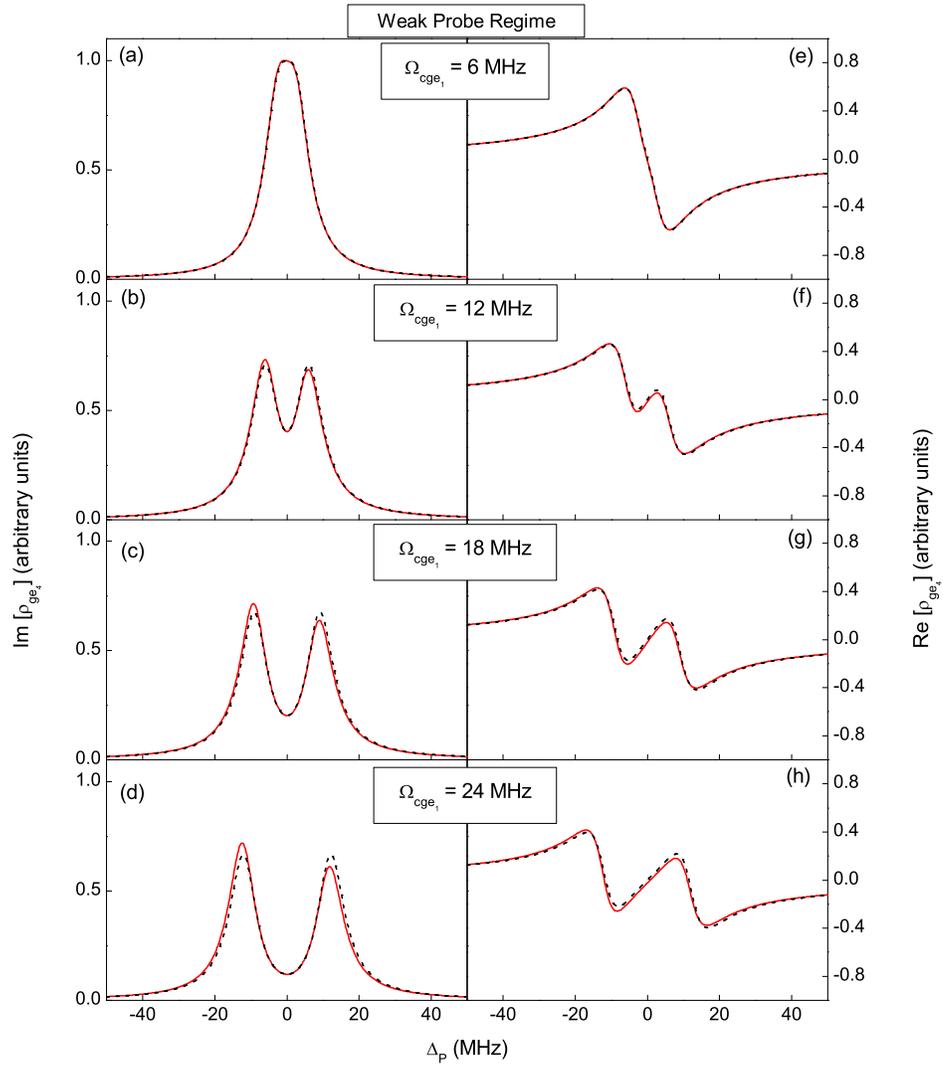}
\caption{The probe absorption and dispersion coefficients Im($\rho_{ge_{4}}$) and Re($\rho_{ge_{4}}$) for atoms as a function of probe detuning ($\Delta_{p}$) in the three-level (black dashed line) and six-level (red solid line) systems for four values of $\Omega_{cge_{1}}$ with $\Delta_{c} = 0$ in the weak probe regime.}
\end{center}
\end{figure}

\begin{figure}[tbh!]
\begin{center}
\includegraphics[width=0.8\columnwidth]{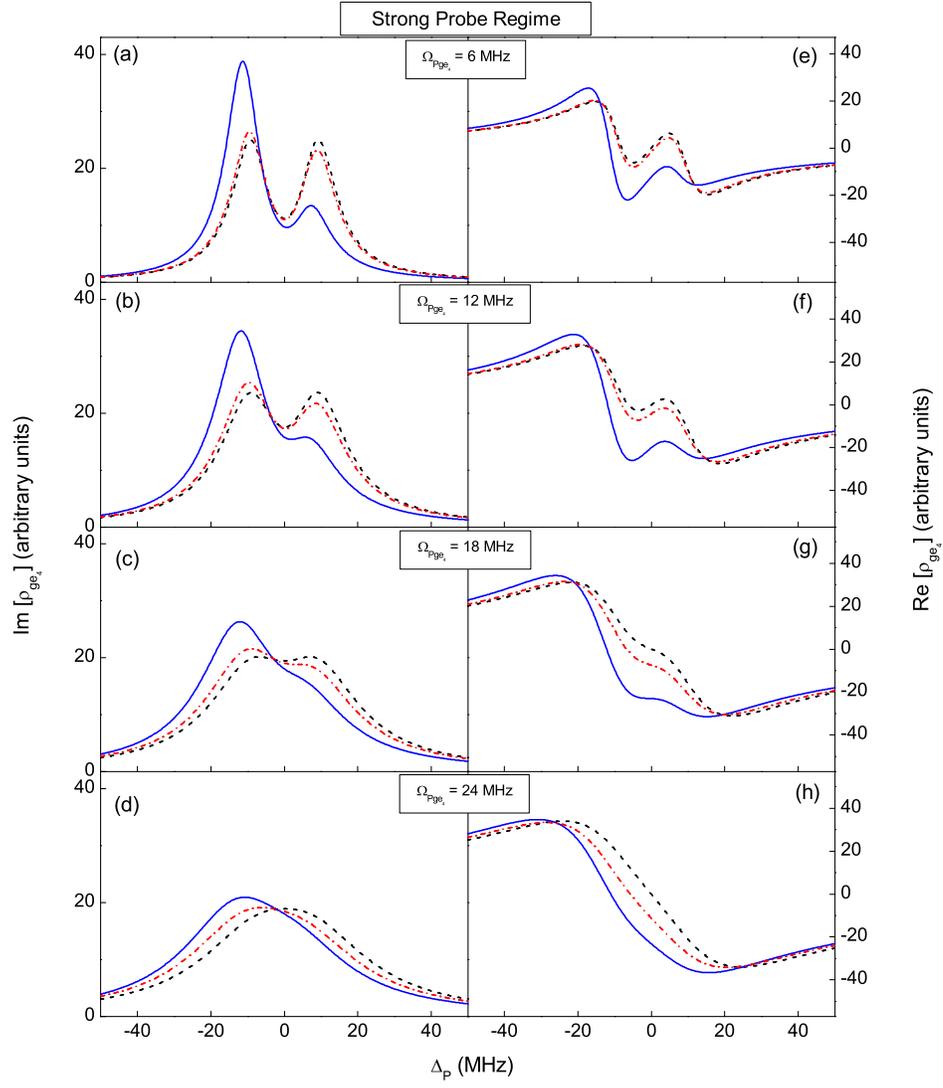}
\caption{The probe absorption and dispersion coefficients Im($\rho_{ge_{4}}$) and Re($\rho_{ge_{4}}$) for atoms as a function of probe detuning ($\Delta_{p}$) for four values of $\Omega_{pge_{4}}$ with $\Delta_{c} = 0$ in the strong probe regime. The red dash-dotted, blue solid and black dashed lines are for the six-, five- and three-level systems, respectively.}
\end{center}
\end{figure}

\begin{figure}[tbh!]
\begin{center}
\includegraphics[width=0.8\columnwidth]{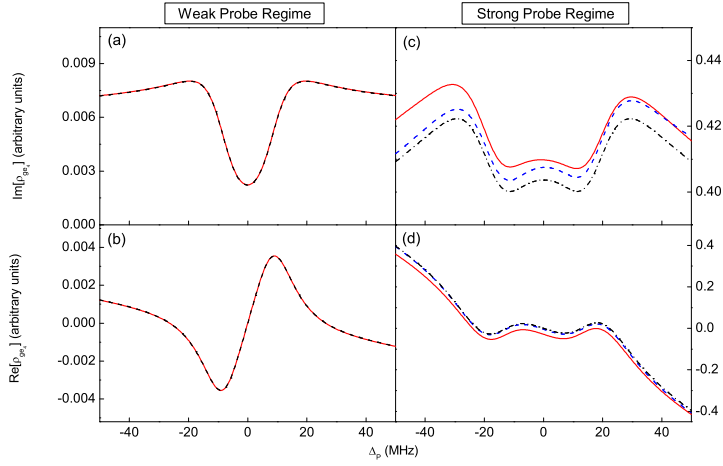}
\caption{The Doppler-broadened probe absorption and dispersion as a function of probe detuning ($\Delta_{p}$) at room temperature 297 K in the weak and strong probe regimes. The red solid, blue and black dashed lines are for the six-, five- and  three-level systems, respectively. In these results, $\Omega_{cge_{1}}=18$ MHz and $\Omega_{pge_{4}}=24$ MHz in weak and strong probe regimes, respectively. All the graphs are normalized by the same factor.}
\end{center}
\end{figure}

\begin{figure}[tbh!]
\begin{center}
\includegraphics[width=0.4\columnwidth]{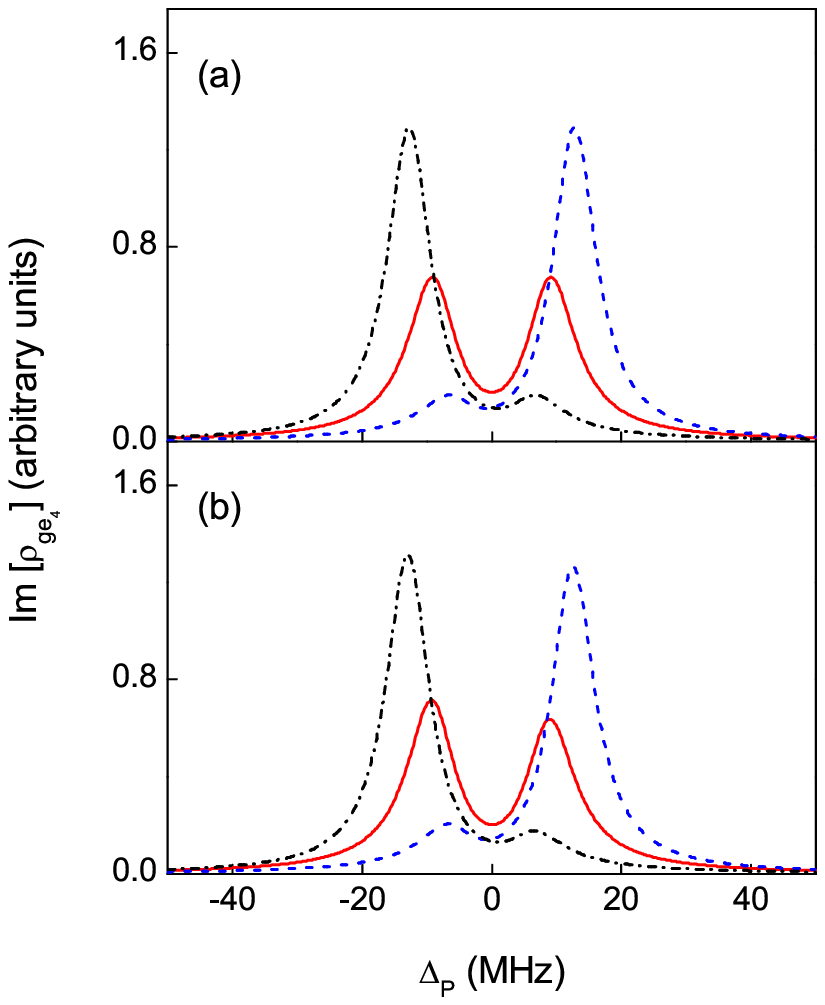}
\caption{Effect of velocity on the probe absorption of (a) three-level and (b) six-level systems for $\Omega_{cge_{1}}=18$ MHz with $\Delta_{c} = 0$ in the weak probe regime. The red solid, blue dashed and black dash-dotted lines are for atoms moving with zero velocity, towards right and left with velocity 5 ms$^{-1}$, respectively.}
\end{center}
\end{figure}

\begin{figure}[tbh!]
\begin{center}
\includegraphics[width=0.3\columnwidth]{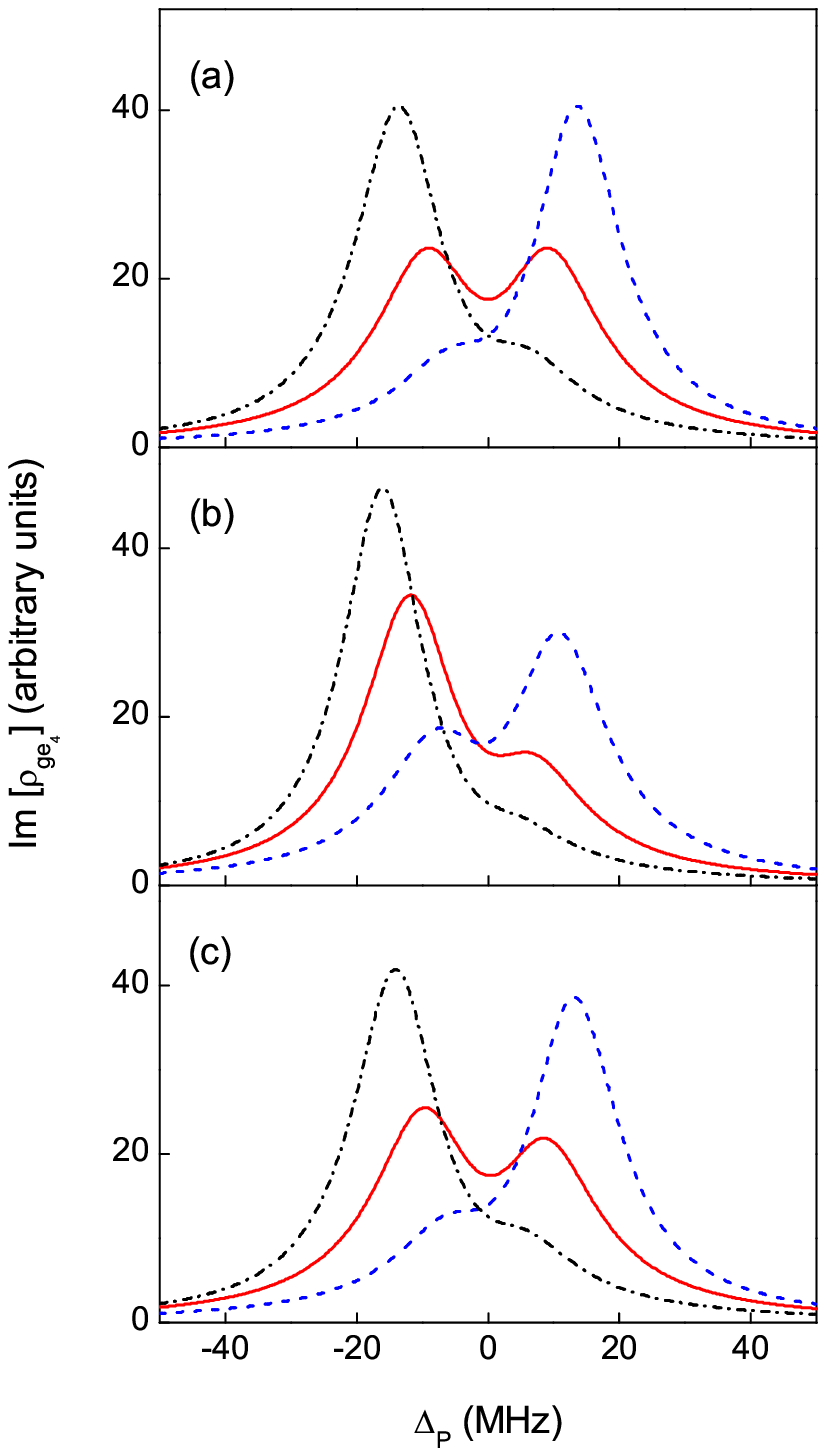}
\caption{Effect of velocity on the probe absorption of (a) three-level (b) five-level and (c)  six-level systems for $\Omega_{cge_{1}}=18$ MHz with $\Delta_{c} = 0$ in the strong probe regime. The red solid, blue dashed and black dash-dotted lines are for atoms moving with zero velocity, towards right and left with velocity 5 ms$^{-1}$, respectively.}
\end{center}
\end{figure}

\begin{figure}[tbh!]
\begin{center}
\includegraphics[width=0.9\columnwidth]{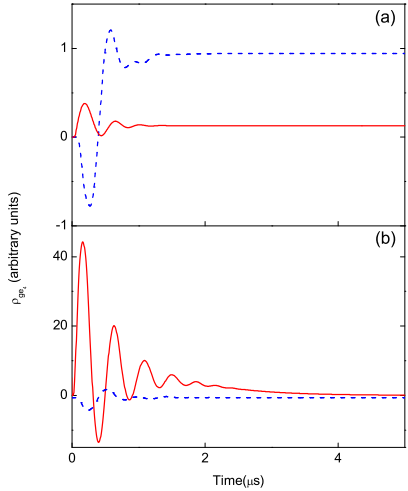}
\caption{ Time evolution of six-level system in the (a) weak probe regime with $\Omega_{cge_{1}}=18$ MHz and (b) strong probe regime with $\Omega_{pge_{4}}=24$ MHz. The red solid and blue dashed curves are for the Im$(\rho_{ge_{4}})$ and Re$(\rho_{ge_{4}})$, respectively.}
\end{center}
\end{figure}

\end{document}